\documentclass[10pt,conference]{IEEEtran}
\IEEEoverridecommandlockouts
\usepackage{array}
\usepackage{amsmath,amsfonts}
\usepackage{algorithmic}
\usepackage{graphicx}
\usepackage{textcomp}
\usepackage{xcolor}
\usepackage{tcolorbox}
\usepackage{listings}
\usepackage{makecell}
\usepackage{adjustbox}
\usepackage{multirow}
\usepackage{svg}
\usepackage{xspace}
\usepackage{amsmath}
\usepackage{subfigure}
\usepackage{amsmath}
\usepackage{booktabs}
\usepackage{color, colortbl}
\usepackage{url}
\usepackage{graphicx}
\usepackage{caption} 
\usepackage{booktabs}
\usepackage{pifont}
\usepackage{dsfont}
\usepackage{floatrow}
\newfloatcommand{capbtabbox}{table}[][\FBwidth]
\usepackage{tabularx}

\newtcolorbox{cbox}[1][]{
   before=\par\smallskip\centering,
   #1
}

\DeclareUnicodeCharacter{2212}{-}
\definecolor{Gray}{gray}{0.9}
\definecolor{green}{RGB}{102,252,102}
\definecolor{ored}{RGB}{255,99,71}
\definecolor{orange}{RGB}{255,165,0}
\definecolor{lightgray}{RGB}{211,211,211}

\usepackage{xcolor}
\usepackage{xcolor,colortbl}
\definecolor{lightgray}{gray}{0.93}
\definecolor{slightgray}{gray}{0.98}
\definecolor{darkgray}{gray}{0.77}
\definecolor{applegreen}{rgb}{0.55, 0.71, 0.0}
\definecolor{chromeyellow}{rgb}{1.0, 0.65, 0.0}
\definecolor{darkpastelgreen}{rgb}{0.01, 0.75, 0.24}

\usepackage[ruled,vlined,linesnumbered]{algorithm2e}

\SetCommentSty{mycommfont}

\usepackage{tikz}
\usetikzlibrary{shapes}
\newcommand*\circled[1]{\tikz[baseline=(char.base)]{
            \node[shape=circle,fill=gray!45, draw,inner sep=1pt] (char) {#1};}}

\makeatletter
    \newcommand{\linebreakand}{%
      \end{@IEEEauthorhalign}
      \hfill\mbox{}\par
      \mbox{}\hfill\begin{@IEEEauthorhalign}
    }
    \makeatother

\newcommand{\tool}{{\textsc{MAdroid}}\xspace}

\newcommand{\chen}[1]{\textcolor{red}{\textbf{Chen}: #1}}
\def\BibTeX{{\rm B\kern-.05em{\sc i\kern-.025em b}\kern-.08em
    T\kern-.1667em\lower.7ex\hbox{E}\kern-.125emX}}
\begin{document}

\title{Agent for User: Testing Multi-User Interactive Features in TikTok}


\author{\IEEEauthorblockN{Sidong Feng}
    \IEEEauthorblockA{\textit{Monash University} \\
    Australia \\
    sidong.feng@monash.edu}
    \and
    

    \IEEEauthorblockN{Changhao Du}
    \IEEEauthorblockA{\textit{Jilin University} \\
    China \\
    chdu22@mails.jlu.edu.cn}
    \and

    \IEEEauthorblockN{Huaxiao Liu}
    \IEEEauthorblockA{\textit{Jilin University} \\
    China \\
    liuhuaxiao@mails.jlu.edu.cn}
    \and

    \IEEEauthorblockN{Qingnan Wang}
    \IEEEauthorblockA{\textit{Jilin University} \\
    China \\
    wangqn23@mails.jlu.edu.cn}
    \linebreakand
    

    \IEEEauthorblockN{Zhengwei Lv}
    \IEEEauthorblockA{\textit{Bytedance} \\
    China \\
    lvzhengwei.m@bytedance.com}
    \and

    \IEEEauthorblockN{Gang Huo}
    \IEEEauthorblockA{\textit{Bytedance} \\
    China \\
    huogang@bytedance.com}
    \and

    \IEEEauthorblockN{Xu Yang}
    \IEEEauthorblockA{\textit{Bytedance} \\
    China \\
    yangxu.swanoofl@bytedance.com}
    \and 
    \IEEEauthorblockN{Chunyang Chen}
    \IEEEauthorblockA{\textit{Technical University of Munich} \\
    Germany \\
    chun-yang.chen@tum.de}
    }


\maketitle

\begin{abstract}
TikTok, a widely-used social media app boasting over a billion monthly active users, requires effective app quality assurance for its intricate features. Feature testing is crucial in achieving this goal. However, the multi-user interactive features within the app, such as live streaming, voice calls, etc., pose significant challenges for developers, who must handle simultaneous device management and user interaction coordination. To address this, we introduce a novel multi-agent approach, powered by the Large Language Models (LLMs), to automate the testing of multi-user interactive app features. In detail, we build a virtual device farm that allocates the necessary number of devices for a given multi-user interactive task.  For each device, we deploy an LLM-based agent that simulates a user, thereby mimicking user interactions to collaboratively automate the testing process. The evaluations on 24 multi-user interactive tasks within the TikTok app, showcase its capability to cover 75\% of tasks with 85.9\% action similarity and offer 87\% time savings for developers. Additionally, we have also integrated our approach into the real-world TikTok testing platform, aiding in the detection of 26 multi-user interactive bugs.
\end{abstract}


\section{Introduction}

\begin{figure}
	\centering
	\includegraphics[width = 0.9\textwidth]{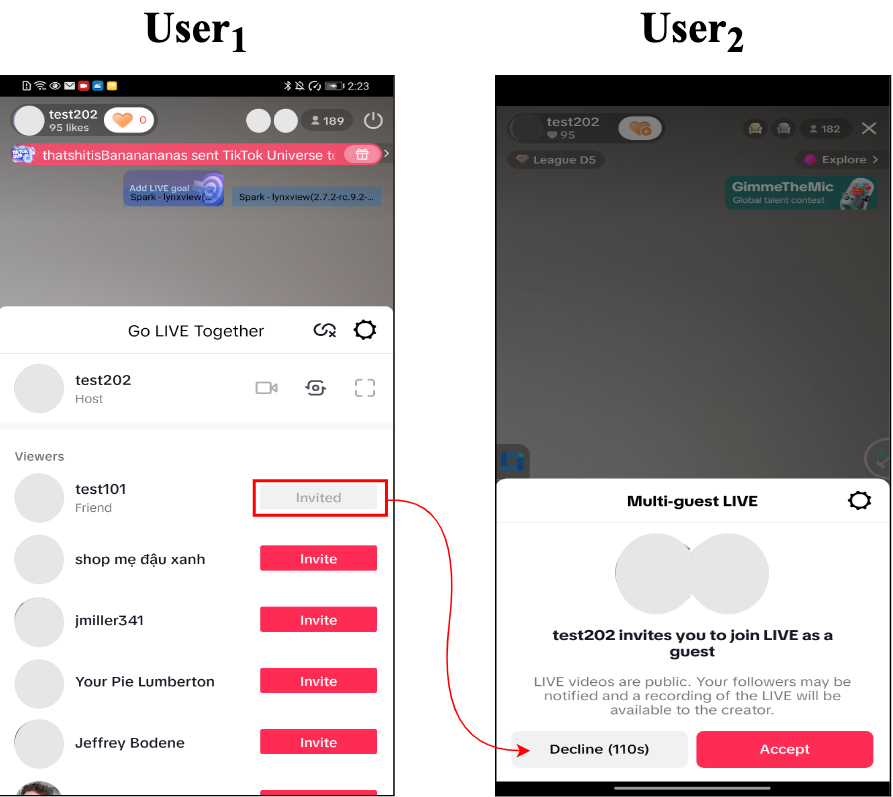}
	\caption{Illustration of LIVE together.}
	\label{fig:example}
\end{figure}

TikTok\footnote{https://www.tiktok.com/en/}, also known as Douyin in China, commands a global user base with over 1.04 billion monthly active users, becoming one of the most popular social media apps in the world.
It offers a diverse set of features, including voice calls, live streaming, and video conferencing, which have transformed the way users engage with each other, facilitating immediate communication, collaboration, and social connection.
With TikTok's rapid expansion and introduction of new features, app quality assurance has become both increasingly crucial and challenging.
Feature testing is an indispensable approach.
This involves developers writing automated scripts tailored to specific functional task objectives, with a focus on covering the testing of the app's primary capabilities.



Nevertheless, testing features within the TikTok app poses distinct challenges.
Our empirical analysis of 7,870 feature testing tasks within TikTok, as discussed in Section~\ref{sec:background}, reveals that 40.31\% of these tasks pertain to the interactions between multiple users. 
For instance, as depicted in Fig.~\ref{fig:example}, a user (User$_{1}$) initiates a multi-guest LIVE session by inviting friends (User$_{2}$) to join a collaborative communication. 
Despite their importance, testing these multi-user interactive features can be exceptionally time-consuming and laborious for developers, due to two major challenges.
First, the task automation test scripts must be adaptable across different devices, a requirement that becomes more complex when it involves the variety of devices used in testing multi-user interactive features.
Second, the sequence of interactions between multiple devices requires precise coordination. 
Any deviation from the intended sequence, such as attempting to accept before an invitation is properly sent, could lead to the feature failing to trigger correctly.
While extensive research~\cite{wang2025empirical,feng2022gifdroid,feng2021auto,feng2023read,feng2022gifdroid1} has been devoted to automated app testing, none have adequately covered the testing of those collaborative and sequential multi-user interactive features across multiple devices.

Recent studies~\cite{kumaresen2020agent,ramirez2021agent} have proposed the use of agent frameworks in the domain of software testing.
In these frameworks, agents are programmed to autonomously perceive and interact with their environment, working together to solve problems.
However, these frameworks are often constrained by hard-coded rules tailored to specific software, which limits their adaptability to apps like TikTok with dynamic GUI contents and flexible user interactions.
In recent years, the rise of advanced Large Language Models (LLMs) has introduced a paradigm shift in this space.
Unlike traditional rule-based agents, LLM-powered agents can adapt to diverse software environments with human-like flexibility.
A notable example is Anthropic’s release of Claude 3.5 Computer Use~\cite{web:claude}, introducing an LLM-powered GUI agent designed for interacting with software interfaces in a more human-like and adaptive manner.

Inspired by the human-like capability of LLMs, we propose a novel multi-agent framework powered by the LLMs to autonomously conduct multi-user interactive feature testing.
This framework employs agents that simulate users interacting with mobile devices, working collaboratively to trigger multi-user interactive features.
Specifically, taking a description of a multi-user interactive task as input, we first build a virtual device farm to support a number of virtual devices to allocate.
Following this setup, we distribute the task to individual agents, each representing a user and paired with a virtual device.
These agents, driven by LLMs, mimic user interactions on their respective devices and autonomously guide their navigation through the GUI screens. 
By operating in a collaborative manner, the agents contribute to the automation of multi-user interactive tasks.

To evaluate the performance of our approach, we first carry out a pilot experiment involving 24 interactive tasks in the TikTok app.
Results demonstrate that our approach achieves a success rate of 75\% and 85.9\% of its actions are the same as the ground truth, significantly outperforming three state-of-the-art baselines.
Additionally, our approach offers a time-saving advantage, reducing the time required for multi-user feature testing by 87\% for professional developers.
Beyond the effectiveness of the approach, we also assess its practicality in the real-world environment.
We integrate our approach with the TikTok testing platform, triggering it whenever new functionalities need testing or a new version is released, accompanied by task descriptions.
In a real-world platform involving the testing of 3,318 multi-user interactive tasks, our approach is instrumental in uncovering 26 interactive bugs, indicating the potential value of our approach in enhancing the practical testing process.

The contributions of this paper are as follows:
\begin{itemize}
    \item We discover the problem of multi-user interactive features within the software.
    \item We propose a novel approach, that harnesses a multi-agent LLMs framework for the autonomous automation of multi-user interactive tasks.
    \item Comprehensive experiments, including the effectiveness and usefulness evaluation of our approach for interactive task automation. Additionally, we incorporate our approach into a real-world testing platform to illustrate its practicality.
\end{itemize}

\section{Background \& Motivation}
\label{sec:background}

In this section, we first define the problem of multi-user interactive features. 
Then, we briefly discuss the practical challenges of testing these interactive features in real-world testing environments, underscoring the necessity for specialized tool support.

\subsection{Definition of Multi-User Interactive Feature}
\label{sec:definition}



\begin{quote}
  \vspace{0.05cm}
  \textbf{Definition}: \textit{Multi-user interactive feature is a component of software that enables \textsuperscript{\circled{\footnotesize{1}}}\underline{multiple users} to \textsuperscript{\circled{\footnotesize{2}}}\underline{interact with each other} in \textsuperscript{\circled{\footnotesize{3}}}\underline{nearly real-time} within a software.} 
  \vspace{0.15cm}
\end{quote}


The definition of multi-user interactive features is dissected into three fundamental elements. 
First, the feature enables users to interact with multiple users through the software.
Note that interactions between a user and the software or the administrator are not categorized under multi-user interactive features. 
An example of this would be the process of activity organization, where an end-user books a reservation and the admin confirms it. 
Such interactions are predominantly administrative and do not embody the peer-to-peer collaborative or communicative intent that is the hallmark of multi-user interactive features.

The second critical aspect of the interactive feature is its fundamental interactivity.
This characteristic is marked by the presence of interactive cues including notifications or pop-up messages, that differ significantly from collaborative functions like text editing.
These cues play a crucial role in ensuring users are aware of other's requests and encouraging them to act or respond.

Third, the interactive feature pertains to a timely nature.
Interactions should occur with minimal delay, though not necessarily instantaneously, to maintain a fluid information exchange and keep users actively engaged. 
For example, when a user sends an invitation, the other should accept the invitation promptly to ensure effective interaction, otherwise, it will expire automatically, e.g., after 110 seconds in Fig.~\ref{fig:example}.

\subsection{Prevalence of Multi-User Interactive Features}
\label{sec:prevalence}


\renewcommand{\arraystretch}{1.22}
\begin{table*}
    \footnotesize
    \tabcolsep=0.14cm
	\centering
	\caption{Multi-user interactive tasks in TikTok. \textit{\#Actions} indicates the total number of actions required for task completion.}
	\label{tab:task_detail}
	\begin{tabular}{l|c|c|c|c} 
	    \hline
	   \bf{Task ID} & \bf{User$_1$} & \bf{User$_2$} & \bf{User$_3$} & \bf{\#Actions} \\
            \hline
            \hline
            1 & invite User$_2$ to a multi-guest LIVE session & accept the invitation & - & 4 \\
            \hline
            2 & send User$_2$ an interactive card in LIVE session & trigger the interactive card & - & 4 \\
            \hline
            3 & send a comment in LIVE session & reply the comment & - & 4 \\
            \hline
            4 & create a face-to-face group chat & join the chat & join the chat & 24 \\
            \hline
            5 & invite User$_2$ to join group chat & accept the invitation & - & 4 \\
            \hline
            6 & send a real-time session invitation in group chat & enter the session via join card & accept the invitation & 7 \\
            \hline
            7 & send a video call in group chat & accept the call & accept the call & 6 \\
            \hline
            8 & send a animation in group chat & resume the animation & - & 4 \\
            \hline
            9 & transfer group chat to User$_2$ & confirm the group transfer & accept the transfer & 9 \\
            \hline
            10 & invite User$_2$ with an animated link and User$_3$ without & join the animated link & join the link & 8 \\
            \hline
            11 & send a video call to User$_2$ by Tool Bar & accept the call & - & 4 \\
            \hline
            12 & send a video call to User$_2$ by Navigation Bar & accept the call & - & 2 \\
            \hline
            13 & send a video call to User$_2$ by Navigation Bar, then hang up & make a call back & - & 4 \\
            \hline
            14 & send a video call to User$_2$ by Navigation Bar & accept the call, and invites User$_3$ to join & accept the call & 6 \\
            \hline
            15 & invite User$_2$ to watch together & accept the invitation & - & 7 \\
            \hline
            16 & invite User$_2$ to watch together & accept the invitation via push notification & - & 6 \\
            \hline
            17 & send an IM join request for play together & accept the invitation & - & 7 \\
            \hline
            18 & send a video feed stream to User$_2$ & pause the stream & - & 5 \\
            \hline
            19 & send a PK invitation to User$_2$ & accept the PK & - & 3 \\
            \hline
            20 & send a connection link request to User$_2$ & accept the link & - & 4 \\
            \hline
            21 & start a payment transfer to User$_2$ & confirm the payment & - & 8 \\
            \hline
            22 & share a live post to User$_2$ & add a comment & - & 8 \\
            \hline
            23 & share a live post to User$_2$ & forward to User$_3$ & add a comment & 14 \\
            \hline
            24 & share a real-time review link to User$_2$ & agree to review & - & 9 \\
            \hline
	\end{tabular}
\end{table*}



To investigate the prevalence of multi-user interactive features in apps, we carry out an empirical study focusing on the TikTok app. 
We collect a dataset for practical feature testing, comprising 7,870 task descriptions sourced from TikTok. 
On average, each description contains 20.64 words and has been identified by 15 internal developers as serving a functionality of the app.
This dataset has been employed in task automation tests throughout 52 app development cycles over a span of one year.

Two authors conduct manual labeling of multi-user interactive features within the dataset.
Note that all annotators possess over two years of experience in mobile app testing and have been daily active TikTok users.
Initially, the annotators independently label whether tasks within the dataset are related to the multi-user interactive feature, drawing on their hands-on experience with the app, without any collaborative discussion.
After the initial labeling, the annotators meet to resolve any minor discrepancies in their annotations.
Any remaining discrepancies will be handed over to a professional internal developer for final adjudication.
To further validate the dataset’s quality, three internal developers are also invited to scrutinize the collected tasks.

We discover that 40.31\% of the task descriptions are related to the multi-user interactive features.
For the brevity of the paper, we showcase the most frequently tested 24 task descriptions in Table~\ref{tab:task_detail}, to serve as our experimental dataset, covering the core functionalities within the TikTok app.
Each task is structured as a user-specific sequence $<$User$_1$: ...; User$_2$: ...; User$_n$: ...$>$.
For example, \textit{Task\#15} is described as ``User$_1$: invite User$_2$ to watch together; User$_2$: accept the invitation''.
Execution of these tasks typically requires the participation of two users, with some tasks requiring the collaboration of up to three users.
We analyze the number of users that are presented in the task description, reflecting the foundational user requirement for each task.
Although this number is indicative, it can be expanded; for instance, a group chat feature may accommodate hundreds of users.

\begin{figure}
	\centering
	\includegraphics[width = 0.98\textwidth]{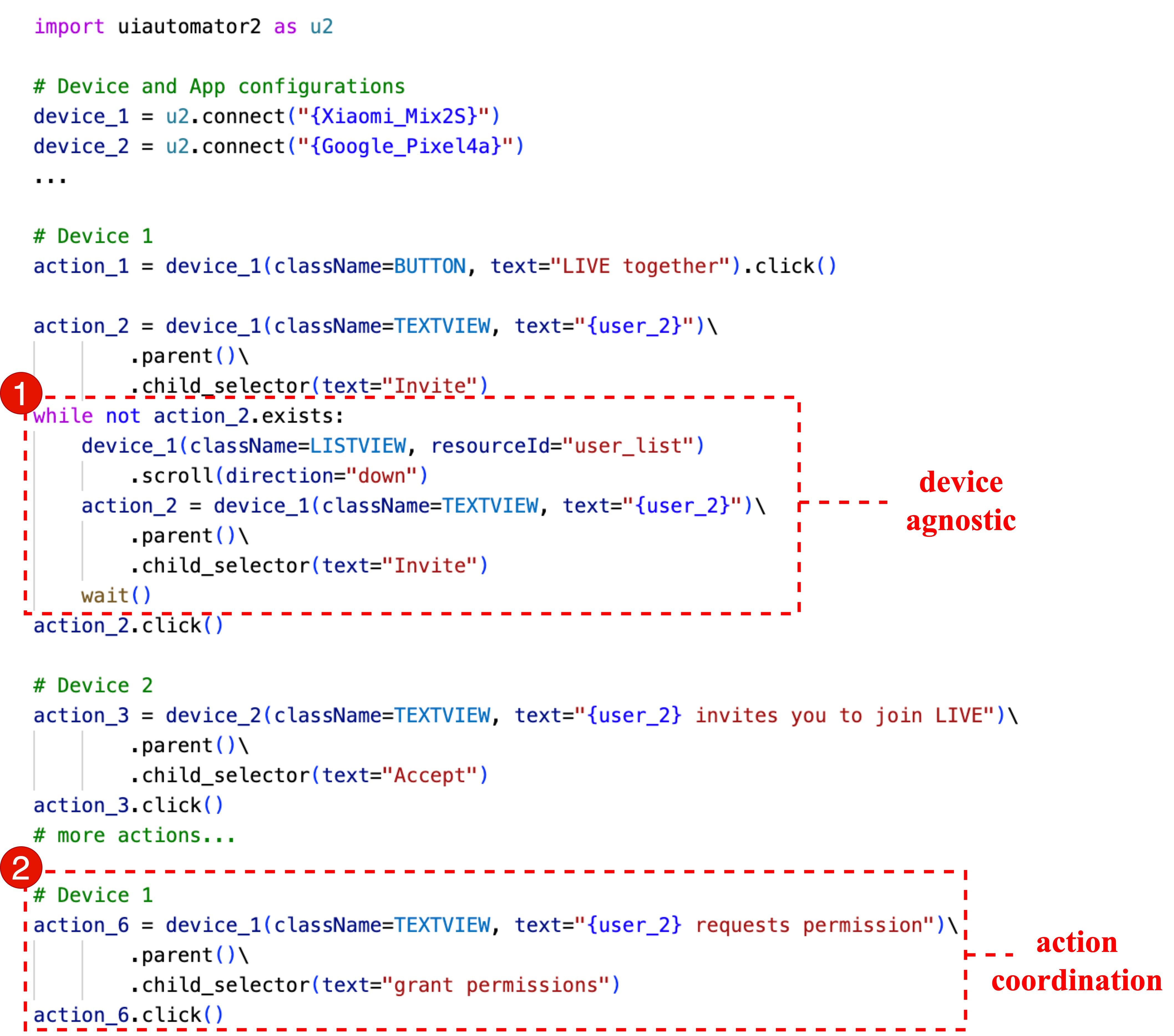}
	\caption{Example of the script for multi-user feature testing.}
	\label{fig:script}
\end{figure}

\subsection{Test of Multi-User Interactive Features}
\label{sec:script}
A common practice for conducting feature testing is to create record-and-replay scripts. 
This includes recording a series of user actions into a script and then automating their execution on devices to monitor any exceptions that are triggered, causing the replaying failure. 
However, writing a practical script for tasks that involve multi-user interaction poses significant challenges.
To better understand these difficulties, we engage in informal discussions with two professional internal developers at TikTok, each with over three years of work experience.
In summary, we identify two primary challenges.

First, the script must be designed to be device-independent in mind, ensuring its functionality across an array of screen sizes, resolutions, and operating systems.
The complexity increases when the script contains the interaction between multiple devices.
For instance, in the scenario of Fig.~\ref{fig:example}, when a user starts a multi-guest LIVE session on an older device with a smaller screen, the GUI presented might not display the target user. 
To address such potential differences in GUIs, the script must be adaptable, including actions like scrolling through the user list, as depicted in Fig.~\ref{fig:script}-1.
This ensures that the interface on the older device functions seamlessly despite any disparities.


Second, the script needs to coordinate actions between users to guarantee that interactions happen in the intended sequence, closely mirroring real-world usage. 
This synchronization challenge is compounded when the script is required to handle interactive features where users perform actions simultaneously or respond to each other's actions.
For example, consider a sequence of interactions in Fig.~\ref{fig:script}-2, where one user requests specific permissions, followed by one who must grant them; these actions must be carried out in a rigid order.
Any mistake in any step will fail the entire testing process.


\subsection{Motivation}
Despite the prevalence of multi-user features in mobile apps — 40.31\% in the TikTok app — the testing of these features often goes overlooked.
Numerous automated testing tools exist~\cite{li2017droidbot,web:monkey,liu2023make}, yet none are equipped to handle the automation of multi-user interactive features.
This is because their underlying frameworks are primarily designed with single-user scenarios in mind, neglecting the complexities of interactions that occur between multiple users.

Agents~\cite{wooldridge1997agent} are software entities designed to perceive and interact with their environment, capable of performing autonomous actions.
Depending on the flexibility of such actions, agents possess the ability to independently initiate tasks and set their own objectives.
Nonetheless, when addressing complex and collaborative problem-solving scenarios, relying solely on a single agent is often insufficient. 
Such scenarios necessitate the adoption of a multi-agent framework, a coalition of multiple model-driven agents that work collaboratively. 
Extending beyond the output of single-agent to provide solutions, the multi-agent framework deploys a team of agents, each bringing specialized knowledge to collectively address sophisticated challenges.
In these setups, several autonomous agents collaborate in activities akin to human teamwork, such as planning, discussion, and decision-making in problem-solving endeavors. 
Essentially, this approach harnesses the communicative strengths of agents, utilizing their generation and response capabilities for interaction.
In our research of testing multi-user interactive features within the software, where interactions between multiple users involve dynamic exchanges of input and feedback, we have embraced the multi-agent paradigm to enhance the collaboration of agents, making it a valuable asset for interactive features.

\section{Approach}
\label{sec:approach}
\begin{figure}
	\centering
	\includegraphics[width = 0.95\textwidth]{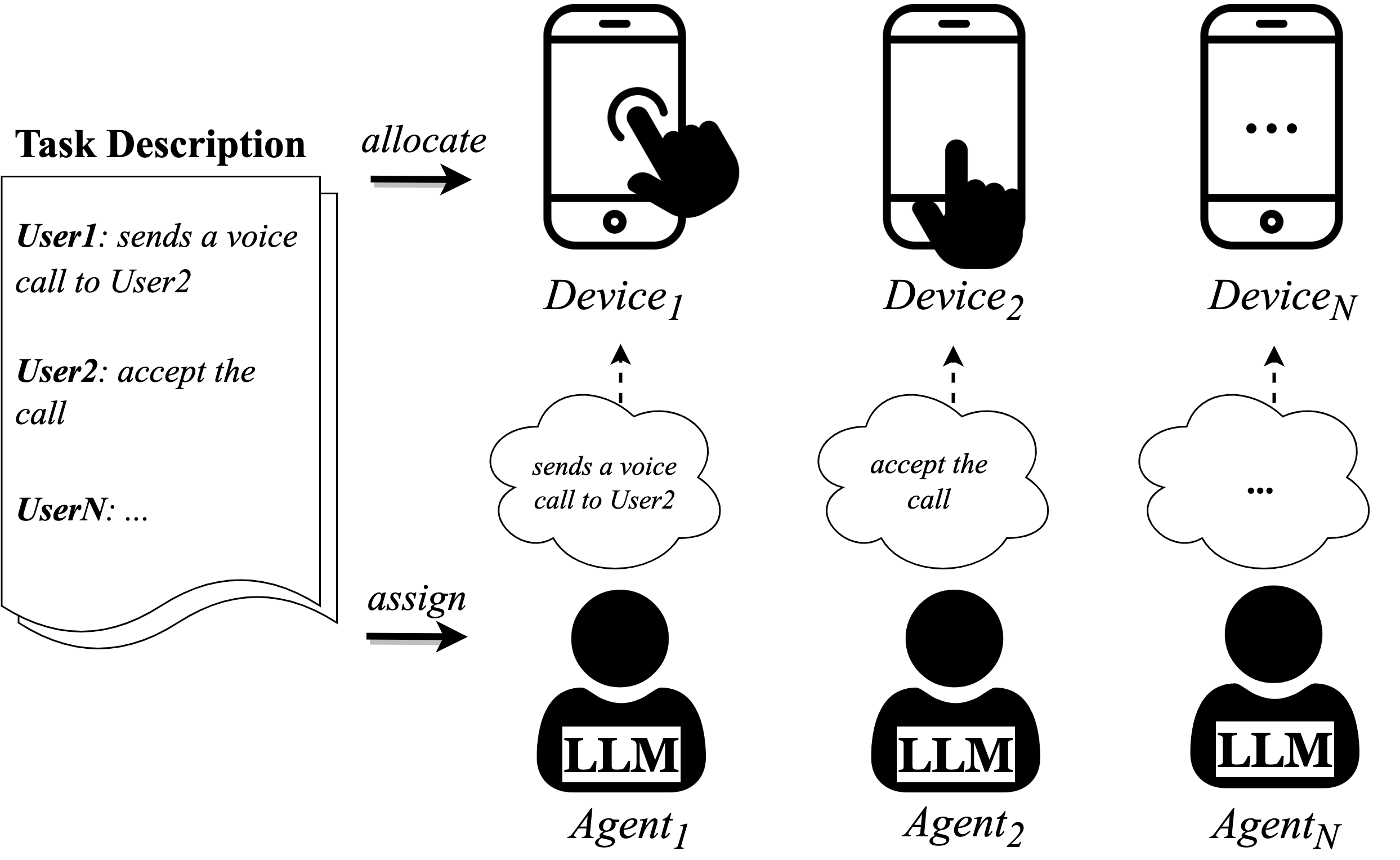}
	\caption{The overview of our approach.}
	\label{fig:overview}
\end{figure}

Given a multi-user interactive task description, we propose an automated approach designed to explore the app to find a sequence of GUI actions to trigger the interactive feature.
The overview of our approach is shown in Fig.~\ref{fig:overview}, which is divided into two main phases: 
(i) the \textit{Device Allocation from Farm} phase, which manages a collection of virtual devices to facilitate device allocation for multi-user scenarios;
and (ii) the \textit{Multi-agent Task Automation} phase, which employs autonomous agents to navigate through dynamic GUIs to accomplish the specified task.

\subsection{Device Allocation from Farm}
\label{sec:phase1}
Since our task involves testing among a number of devices, the initial stage of our approach is to organize and allocate device resources.
It is standard to utilize cloud-based mobile app testing services that offer a device farm containing numerous physical devices for testing purposes, such as AWS Device Farm~\cite{web:aws} and Google Firebase Test Lab~\cite{web:firebase}.
However, considering the significant ongoing costs associated with mobile app testing in the industry, we propose a virtual device farm.
This would involve emulating mobile devices on standard servers through the use of virtualization technologies, presenting a more cost-effective solution for mobile app testing.

\subsubsection{Virtual Device Farm}
We build a virtual device farm as a digital twin of the physical device farm.
Within this virtual environment, we create 5,918 virtual devices using Android emulators distributed across 395 ARM-based commodity servers. 
Each virtual device is designed to mirror a corresponding physical device.
The servers are configured with a 64-core ARM v8.2 CPU operating at 2.6 GHz, equipped with 128 GB of DRAM, 1 TB of NVMe SSD storage, and networked with a 1 Gbps NIC bandwidth.
To ensure the servers can handle the load, a prior empirical study~\cite{lin2023virtual} on load testing confirms the 395 servers can adequately support all 5,918 virtual devices in simultaneous operation.

We opt for ARM servers over x86 servers as the foundational infrastructure for our virtual device farms due to two primary reasons.
First, ARM servers offer a more cost-effective solution, being approximately 23\% less expensive than x86 servers when comparing systems with the same number of CPU cores and similar memory/storage capacities~\cite{web:cloud}.
Second, the native compatibility between ARM servers and Android phones—which all utilize ARM CPUs—eliminates the need for dynamic binary translation (DBT) to convert the ARM instructions of Android apps to the x86 instruction set used by server CPUs~\cite{probst2002dynamic,web:arm}.
This compatibility is especially beneficial considering the rarity of x86-native libraries in mobile apps.

Each virtual device is set up to mirror its physical counterpart in terms of CPU cores, memory and storage capacity, and display resolution.
To simulate CPU locality, we assign each virtual core to a corresponding physical core.
As the multi-user interactive feature relies on network connectivity, we diverge from the typical wired Ethernet approach in virtual devices. 
We tailor the virtual network configuration to match the physical device's network setup, such as WiFiManager~\cite{web:wifimanager} for WiFi networks and DcTracker~\cite{web:dctracker} for cellular networks.

To address the potential variations in apps that may arise across different smartphone vendors, we ensure that each virtual device is equipped with the appropriate vendor-specific app service platforms.
This includes installing a suite of vendor apps and SDKs, such as Google Mobile Services (GMS) for devices that support them and Huawei Mobile Services (HMS) for relevant Huawei devices, to align with the software ecosystem of the physical device it represents. 

\subsubsection{Device Allocation}
To assess the required number of interactive devices based on the given task description, we commence by pinpointing the total number of users involved.
This is achieved by employing a regular expression (regex) pattern to dissect the task description and extract user identifiers.
Specifically, we formulate a regex pattern like ``\texttt{User$_{[1-9]}$}'', where ``\texttt{User}'' is the prefix denoting each user, followed by a number ``\texttt{[1-9]}'' representing the unique index.
By applying this pattern, we determine the number of distinct users, which corresponds to the quantity of required interactive devices that we need to allocate from our virtual device farm.
Note that we select these virtual devices from our pool randomly to ensure a general testing environment.


\subsection{Multi-agent Task Automation}
\label{sec:phase2}
After finalizing the allocation of devices, we proceed by distributing tasks and setting up the order in which they will be initiated. 
Subsequently, we deploy an individual agent onto each device, simulating a user, to autonomously navigate the GUI screen and accomplish the assigned task.

\subsubsection{Task Assignment}
Given the task description, we break it down into discrete tasks that can be assigned across the interactive devices. 
This process involves a potential plan of the task to identify discrete components that can be independently executed by different devices.
We employ rule-based pattern recognition to identify tasks associated with specific ``\texttt{User$_{X}$}'', where ``\texttt{X}'' represents the user index.
For instance, in Fig.~\ref{fig:overview}, take the task ``User$_{1}$: sends a voice call to User$_{2}$; User$_{2}$: accept the call''.
This would be segmented into separate tasks such as ``sends a voice call to User$_{2}$'' and ``accept the call'' for User$_{1}$ and User$_{2}$, respectively.

When dealing with interactions between devices, the sequence in which they undertake their respective tasks is crucial. 
Taking the voice call scenario as an example, the correct sequence would have User$_1$ starting the call; this action must logically come before User$_2$ can accept the call. 
According to a pilot study of task descriptions within the industry, we determine that the initial action would invariably originate from User$_1$.
Note that we only determine the first action, rather than mapping out the full interaction sequence, since subsequent actions may involve inter-device interaction.


\begin{figure}
	\centering
	\includegraphics[width = 0.975\textwidth]{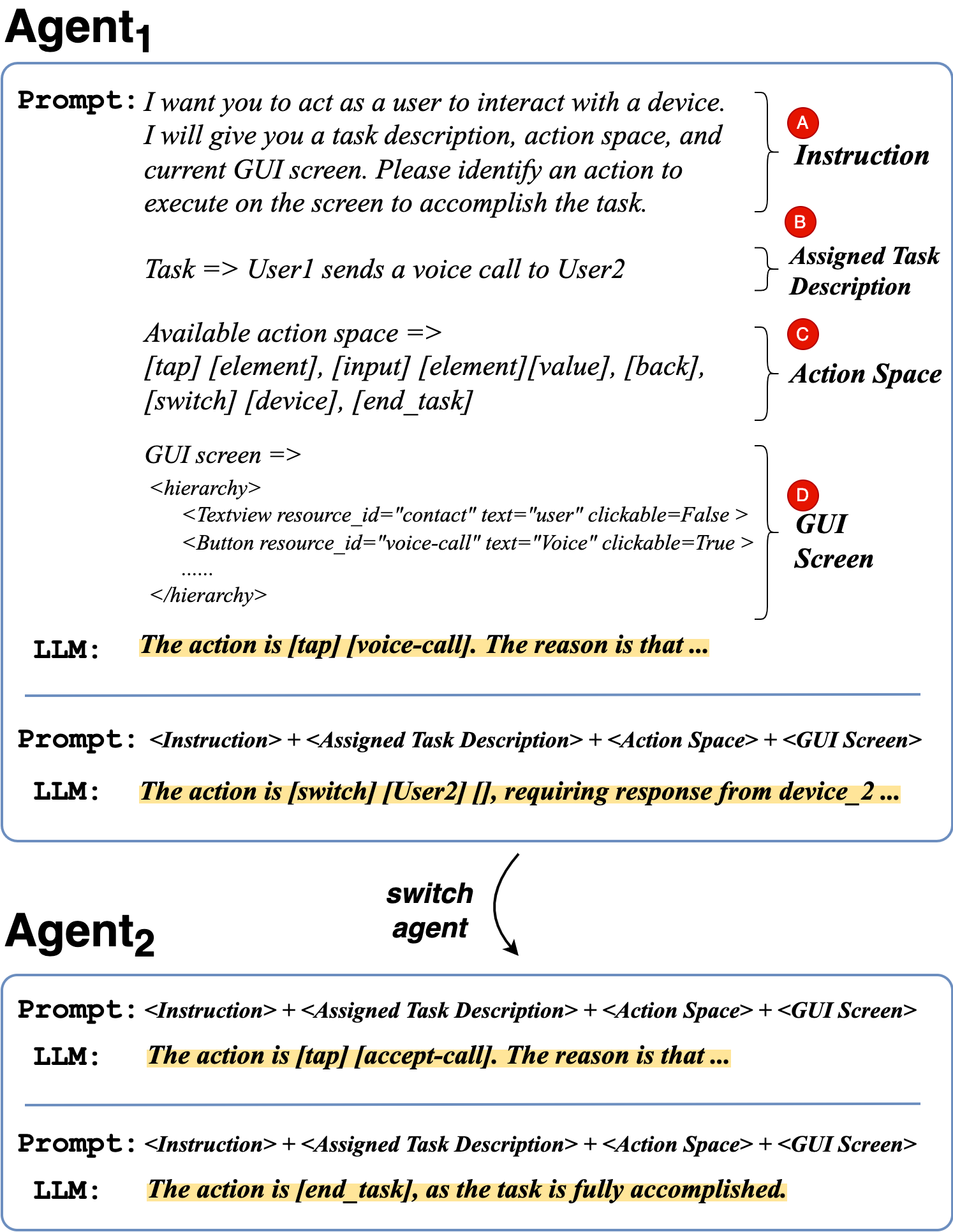}
	\caption{The example of prompting agent.}
	\label{fig:operator}
\end{figure}

\subsubsection{User agent}
To mimic the diversity of users on GUI navigation to work collaboratively to trigger the multi-user interactive feature, we aim to deploy the agent paradigm for the user. 
Inspired by the advance of Large Language Models (LLMs) that exhibit impressive capabilities in communication that closely resemble humans in response to various prompts, we adopt the LLMs as the backbone of the agent paradigm.
However, employing LLMs to operate on the devices presents two primary challenges:
First, LLMs are not inherently equipped for GUI interaction, which may result in a lack of operational knowledge pertaining to the devices.
To address this, we supply the LLMs with a list of actions and their corresponding primitives to facilitate interaction with the GUI screen.
Second, GUIs are typically rich in spatial information, embodying a structural complexity that LLMs might not readily comprehend from visual GUI screenshots~\cite{yang2023dawn}. 
To overcome this, we introduce a heuristic approach that converts the GUIs into domain-specific representations, which are more accessible to the LLMs' understanding. 


\textbf{Action space.}
We have delineated five core actions, which encompass three standard actions, including tap, input, and back, as well as two task-specific actions, including switch\_user, and end\_task.
Although additional customized actions like pinch and multi-finger gestures exist, they are less prevalent in industry testing.

The representation of each action is contingent on its specific context, necessitating distinct primitives to encapsulate the involved entities. 
For instance, the ``tap'' action mandates a target element within the GUI, like a button. 
Therefore, we express this as [tap] [element]. 
When it comes to the ``input'' action, it entails inputting a designated value into a field, and we structure this as [input] [element] [value]. 
Additionally, we accommodate a system action that does not stem directly from the current GUI screen: ``back'', which allows for navigation to a previously visited screen.

We extend our support to actions specifically designed for interactions. 
Upon approaching the potential completion of its assigned task, one agent sends a signal to switch control to another agent to continue the work in sequence. 
To represent this interactive communication action, we employ the notation [switch] [user]. 
Additionally, we have defined an [end\_task] action, which serves as an indicator for models to identify the completion of the task.

\textbf{GUI screen representation.}
To guide LLMs in navigating GUIs, it is essential to dynamically present the contextual information of the current GUI screen, which encompasses elements such as text, type, images, etc. 
For this purpose, we represent a mobile GUI's structure in XML view hierarchy format utilizing Accessibility Service~\cite{web:accessibility}.
These XML representations map out the hierarchical relationships among GUI elements and include various attributes, like class names, text, descriptions, and clickability, that convey the functionality and visual characteristics of the GUI elements. 
However, these attributes can sometimes contain extraneous details, such as text color, shadow color, and input method editor (IME) options, which may complicate the LLMs' ability to interpret the screen.

To address this, we first simplify the GUI elements, by removing non-essential attributes to concentrate on those that are critical for comprehension of the GUI.
In order to ascertain which attributes are indispensable, we conducted an in-depth review of the Android documentation~\cite{web:accessibility} and previous research~\cite{feng2023prompting,feng2024mud}, pinpointing GUI attributes that are semantically significant for interaction. 
Consequently, we opt to use a subset of properties from the element.

\begin{itemize}
    \item \textit{resource\_id}: describes the unique resource id of the element, depicting the referenced resource.
    \item \textit{class}: describes the native GUI element type such as TextView and Button.
    \item \textit{clickable}: describes the interactivity of the element.
    \item \textit{text}: describes the text of the element.
    \item \textit{content\_desc}: conveys the content description of the visual element such as ImageView.
\end{itemize}

Next, we refine the screen layout by discarding the empty layout container. 
For this purpose, we employ a depth-first search algorithm to navigate through the view hierarchy tree and identify nested layouts. 
We systematically iterate over each node, beginning with the root, and eliminate any layouts that consist of a single node, subsequently proceeding to its child node. 
This method not only guarantees a cleaner depiction of the GUI screens but also markedly decreases the number of tokens, that the LLMs will process.

\textbf{Inferring actions.}
Considering the assigned task, the range of possible actions, and the current GUI screen, we prompt the LLMs to infer a single viable action to execute on the screen, thereby advancing one step toward task completion. 
An illustrative example of such prompting is provided in Fig.~\ref{fig:operator}. 
Due to the robustness of LLMs, the prompt sentence does not need to adhere strictly to grammar rules~\cite{brown2020language}.
For instance, given the assigned task to send a voice call in Fig.~\ref{fig:operator}, the output of the action to be performed on the current GUI screen would be [tap] [voice-call].
The agent then executes the suggested actions on the device by using the Android Debug Bridge (ADB)~\cite{web:adb}.
This cycle of prompting LLMs for a plausible action is repeated until the LLMs recognize the need to interact with a different agent, at which point infer a transition command by [switch] in Fig.~\ref{fig:operator}.
When an agent concludes that the assigned task has been fully executed without the need for other agent interaction, it generates a signal indicating task completion (i.e., [end task]).




\subsection{Implementation}
Our approach has been developed as a fully automated tool for triggering multi-user interactive features within the apps given a task description. 
In a preliminary pilot study, we utilized the pre-trained ChatGPT model from OpenAI~\cite{web:chatgpt}.
The foundational model for ChatGPT is the gpt-4 model, which is currently among the most advanced LLMs available.
It's important to note that we assign separate LLM instances of each agent in Section~\ref{sec:phase2}) by each running on different threads.
This makes the agent focus on its own GUI screens, preventing any GUI navigation confusion among the agents.
To automatically parse the LLMs' output, we provide structured formatting instructions that allow for straightforward interpretation using regular expressions (i.e., []). 
We use Android UIAutomator~\cite{web:uiautomator} for extracting the GUI XML view hierarchy, and Android Debug Bridge (ADB)~\cite{web:adb} to carry out the actions to devices.


\section{Evaluation}
\label{sec:evaluation}
In this section, we describe the procedure we used to evaluate our approach in terms of its performance.

\begin{itemize}
    \item \textbf{RQ1:} How effective is our approach in automating multi-user interactive tasks?
    \item \textbf{RQ2:} How useful is our approach in automating multi-user interactive tasks?
    \item \textbf{RQ3:} What is the practical usage of our approach in real-world multi-user interactive feature testing?
\end{itemize}

For \textbf{RQ1}, we carry out experiments to check the effectiveness of our approach in automating multi-user interactive tasks, compared with three state-of-the-art methods.
For \textbf{RQ2}, we examine the usefulness of our approach in assisting developers with testing interactive features.
For \textbf{RQ3}, we assess the practicality of our approach in facilitating the identification of interactive bugs in real-world development settings.

\subsection{RQ1: Effectiveness of our approach}
\label{sec:rq1}
\textbf{Experimental Setup.}
To answer RQ1, we evaluate the ability of our approach to effectively automate multi-user interactive tasks, given only a high-level task objective description.
We utilize the experimental dataset collected in Section~\ref{sec:prevalence}, including 24 multi-user interactive tasks in the Tiktok app.
The details of the task description can be seen in Table~\ref{tab:task_detail}.
To further construct the ground truth of the execution trace for these tasks, two authors are then to navigate through the TikTok app to collect the corresponding GUI execution traces.
In total, we obtain 24 multi-user interactive tasks, with an average of 6.7 actions required per device to complete each task.
Note that, all the tasks are freshly identified and labeled by human annotators specifically for this study, mitigating the potential bias for data leakage that could arise from the use of LLMs.

\textbf{Metrics.}
We employ two evaluation metrics to evaluate the performance of our approach.
The first is the \textit{success rate}, a commonly used metric in task automation that measures the ability of the approach to successfully complete an interactive task within an app.
A higher success rate indicates a more reliable and effective approach.
However, the success rate (either 1 or 0) is too strict without indicating the fine-grained performance of different approaches. 
Hence, we also evaluate the accuracy of the action sequences by comparing them with the ground-truth trace.
Since actions are executed in a specific order, a single accuracy metric may not be sufficient. 
Therefore, we adopt an \textit{action similarity} metric from prior research~\cite{feng2022gifdroid,feng2022gifdroid1,feng2023video2action}, which involves first computing the Longest Common Subsequence (LCS) between the inferred action sequence and the ground-truth trace.
Then, we calculate the similarity score using the formula $\frac{2 \times M}{T}$, where $M$ is the length of the LCS and $T$ is the sum of the lengths of both sequences.
This similarity score ranges from 0\% to 100\% when expressed as a percentage, with higher values indicating greater alignment with the ground truth. 
A perfect match between the generated trace and the ground truth would yield a similarity value of 100\%.

\textbf{Baselines.}
We set up three state-of-the-art methods as baselines for comparison with our approach.
These methods include one task-driven (AdbGPT) and two random-based (Monkey, Humanoid), all commonly utilized in automated app testing.
\textit{AdbGPT}~\cite{feng2024prompting} employs the recent advancements of prompting engineering strategies such as in-context learning and chain-of-thought reasoning for automating bug tasks.
For a fair comparison, we add the prompt for achieving switch user action (i.e., [switch] [user]) and maintain the same configuration of LLMs to avoid any evaluation bias.

\textit{Monkey}~\cite{web:monkey} is a well-known automated testing tool that generates random GUI actions to explore app features. \textit{Humanoid}~\cite{li2019humanoid} leverages a deep neural network that has been trained on real-world human interactions to simulate common feature tests.
For a fair comparison, we launch a corresponding number of devices and run the automated testing tools (i.e., Monkey and Humanoid) on apps in hopes of triggering and achieving the task.
To address potential biases of testing coverage due to time limitations, we have designated a 2-hour testing period for each tool.

\renewcommand{\arraystretch}{1.2}
\begin{table}
    \small
	\centering
	\caption{Performance comparison of state-of-the-art.}
	\label{tab:rq1_performance}
	\begin{tabular}{l||c|c} 
	    \hline
	   \bf{Method} & \bf{Success Rate} & \bf{ Action Similarity}\\
	    \hline
            AdbGPT~\cite{feng2024prompting} & 29.2\% & 55.6\% \\
            \hline
            Monkey~\cite{web:monkey} & 4.2\% & - \\
            \hline
            Humanoid~\cite{li2019humanoid} & 8.3\% & - \\
            \hline
            Our approach & \textbf{75.0\%} & \textbf{85.9\%} \\
            \hline
	\end{tabular}
\end{table}

\begin{figure}
	\centering
	\includegraphics[width = 0.95\textwidth]{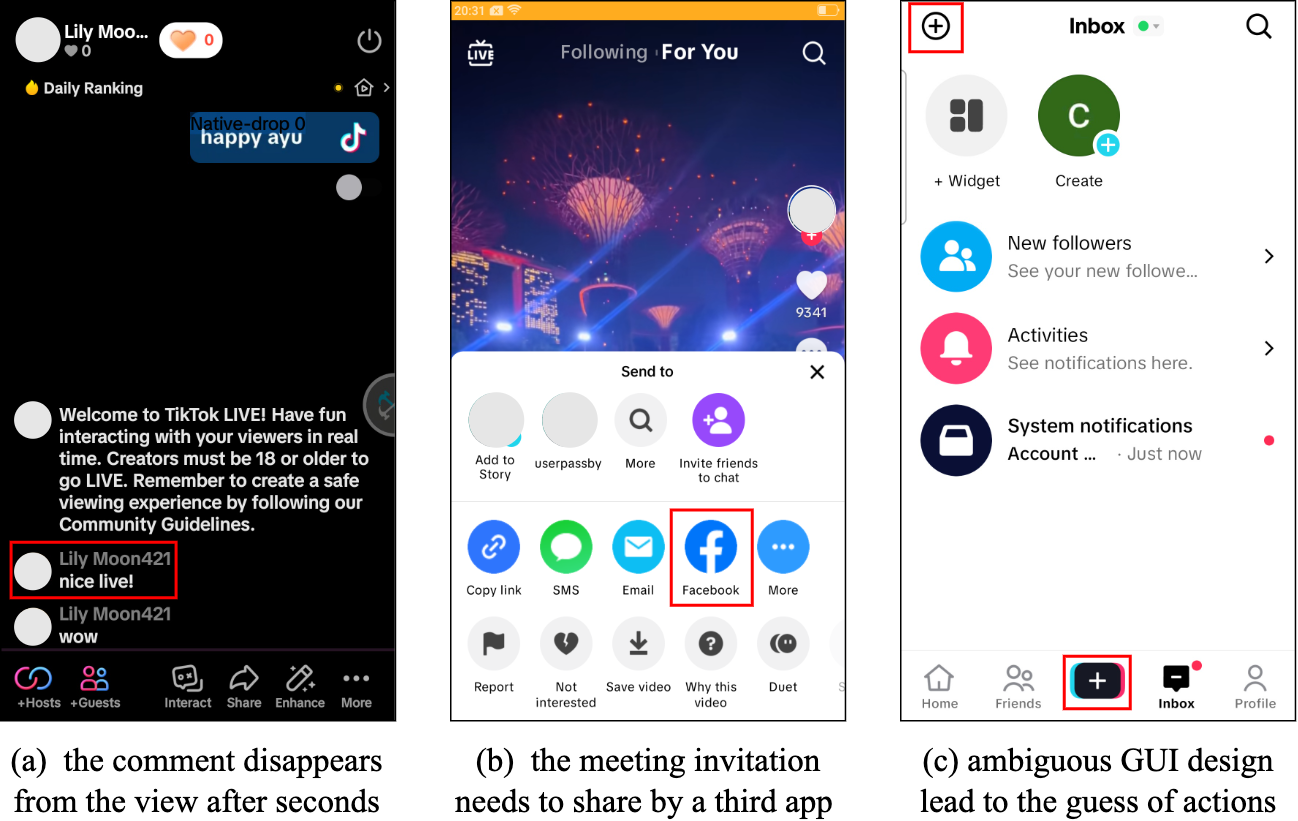}
	\caption{Failure examples of our approach. }
	\label{fig:rq1-2}
\end{figure}


\textbf{Results.}
Table~\ref{tab:rq1_performance} presents a performance comparison of our approach against that of the state-of-the-art baselines. 
Note that all the results are conducted across three runs to address the potential variability of the LLMs.
Our approach achieves an average action similarity of 85.9\%, successfully completing 75\% of multi-user interactive tasks.
The performance of our approach significantly outperforms other baselines, i.e., an improvement of 45.8\% in success rate and 30.3\% in action similarity even compared with the best baseline (AdbGPT).
This is because AdbGPT is prone to making incorrect LLM inferences about the actions to be performed on the screen.
This discrepancy may stem from the fact that its goals are slightly different from ours, focusing on reproducing the bugs through detailed step-by-step descriptions rather than triggering the specific multi-user interactive feature based on a high-level objective.
Such a difference in focus can lead to variations in the prompts, potentially causing a decline in performance.

We observe that the state-of-the-art automated app testing methods (Monkey and Humanoid) do not work well in our task, i.e., only successfully achieve 4.2\% and 8.3\% of the multi-user interactive tasks, respectively. 
Note that since the random-based methods (Monkey and Humanoid) are not tailored to execute specific tasks, we do not calculate action similarity for them; their approach is to randomly navigate the app in hopes of accomplishing the tasks. 
This indicates that random exploration can incidentally test some level of multi-user interactive features, particularly for simpler interactions, such as accepting a call on one device when another device initiates it. 
However, this approach falls short for more complex multi-user interactive tasks, such as joining meetings with code, which demands that the device navigates to the correct GUI screen to enter the specific code.

Albeit the good performance of our approach, we still fail in some multi-user interactive tasks.
We manually examined these failure cases and identified three common causes.
The primary cause of failure is related to timing constraints within the interactions. 
Some multi-user interactive features may require a very quick response (i.e., $<$ 2 seconds).
For instance, the \textit{Task\#3} of responding to the scrolling comments in Fig.~\ref{fig:rq1-2}(a); by the time the LLMs have completed its inference, the comment may have already disappeared from the view.
We are optimistic that this challenge may be mitigated by employing more advanced LLMs or by utilizing local LLMs for quicker inference, thereby reducing delays and improving the success rate of task automation.
Second, multi-user interactive features can span across multiple apps.
For instance, sharing a video link via an external app in \textit{Task\#18} in Fig.~\ref{fig:rq1-2}(b), requires interaction with not just the current app but also the third-party app, e.g., Facebook.
Our approach is limited by the extent to which it requires these back-and-forth interactions across different apps.
Third, ambiguous GUI designs can lead to confusion for the LLMs.
For instance, as depicted in Fig.~\ref{fig:rq1-2}(c), to invite to join a chat in \textit{Task\#5}, the presence of two ``+'' buttons on the GUI screen causes the approach to mistakenly select the incorrect one, thereby struggling to break free from that erroneous exploration path.

\subsection{RQ2: Usefulness of our approach}
\label{sec:rq2}
\textbf{Experimental Setup.}
To answer RQ2, we assess the perceived usefulness of our approach in assisting developers with testing multi-user interactive tasks.
We invite the expertise of two internal developers from TikTok, each with over three years of experience in app testing, to participate in the experiment.
Each developer is asked to write an industrial-level task automation script to trigger the 24 multi-user interactive features in Table~\ref{tab:task_detail}.
To mitigate the threat of user distraction, we conduct the experiment in a quiet room individually without mutual discussion. 
To ensure an equitable comparison with our automated approach, we record the time they spent from the moment they began writing a task automation script to the point where the script was fully automated on devices.



\begin{figure}
	\centering
	\includegraphics[width = 0.83\textwidth]{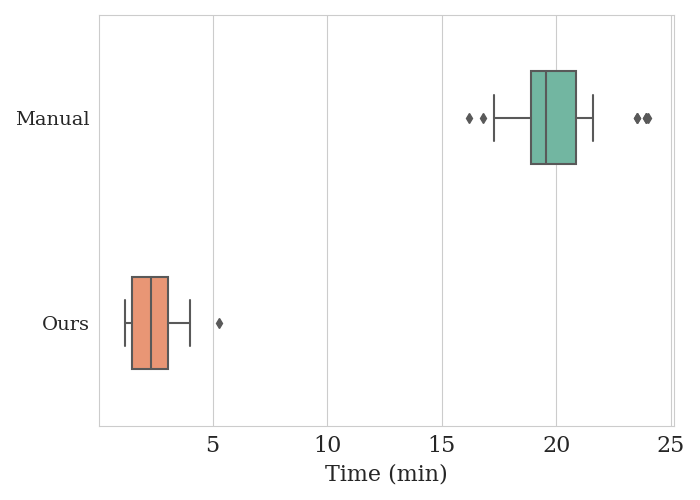}
	\caption{Comparison of time performance}
	\label{fig:rq2}
\end{figure}

\textbf{Results.}
Fig.~\ref{fig:rq2} presents a comparison of the time required for manual human effort vs our automated approach in multi-user interactive task automation. 
On average, our approach requires only 2.46 minutes to automate a task, which is substantially faster than the manual alternative, e.g., $p < 0.05$ in Mann-Whitney U test~\cite{fay2010wilcoxon} (specifically designed for small sample comparison).
In addition, our approach is fully automated and can be executed offline.
In contrast, we have observed that it takes experienced developers an average of 20.02 minutes, and sometimes upwards of 24.0 minutes, to manually write scripts for multi-user interactive task automation. 
This longer duration is primarily due to the complexities of creating device-agnostic scripts and synchronizing actions across multiple devices, as discussed in Section~\ref{sec:script}.
It is worth mentioning that our experiment represents an optimal scenario for manual scripting; the process could be even more daunting for less experienced, junior app developers.
As a result, our approach can expedite 87\% the testing process on the multi-interactive features, thus saving a considerable amount of time during extensive industrial testing which may involve hundreds or thousands of tasks. 


\subsection{RQ3: Practicality of our approach}
\label{sec:rq3}
\textbf{Industrial Usage.} 
In collaboration with TikTok, we have successfully integrated our approach into their testing framework Fastbot~\cite{lv2022fastbot2}.
Our approach is integrated into their internal task automation process and is activated in two ways: 1) when a new feature is proposed given a new task description, and 2) when a new app version is released, the automation of prior tasks for regression testing.
In addition, the primary objective of task automation is bug detection, therefore, our approach is also coupled with their internal bug detection systems.
Specifically, we employ heuristics to monitor event logs for the automatic identification and reporting of crash-related bugs.
For non-crash bugs, we conduct manual monitoring of instances where task automation does not succeed, utilizing the internal developer-friendly system to identify functional bugs.
We set up the number of bugs as the evaluation metric to assess the practicality of our approach.

\begin{figure}
	\centering
	\includegraphics[width = 0.975\textwidth]{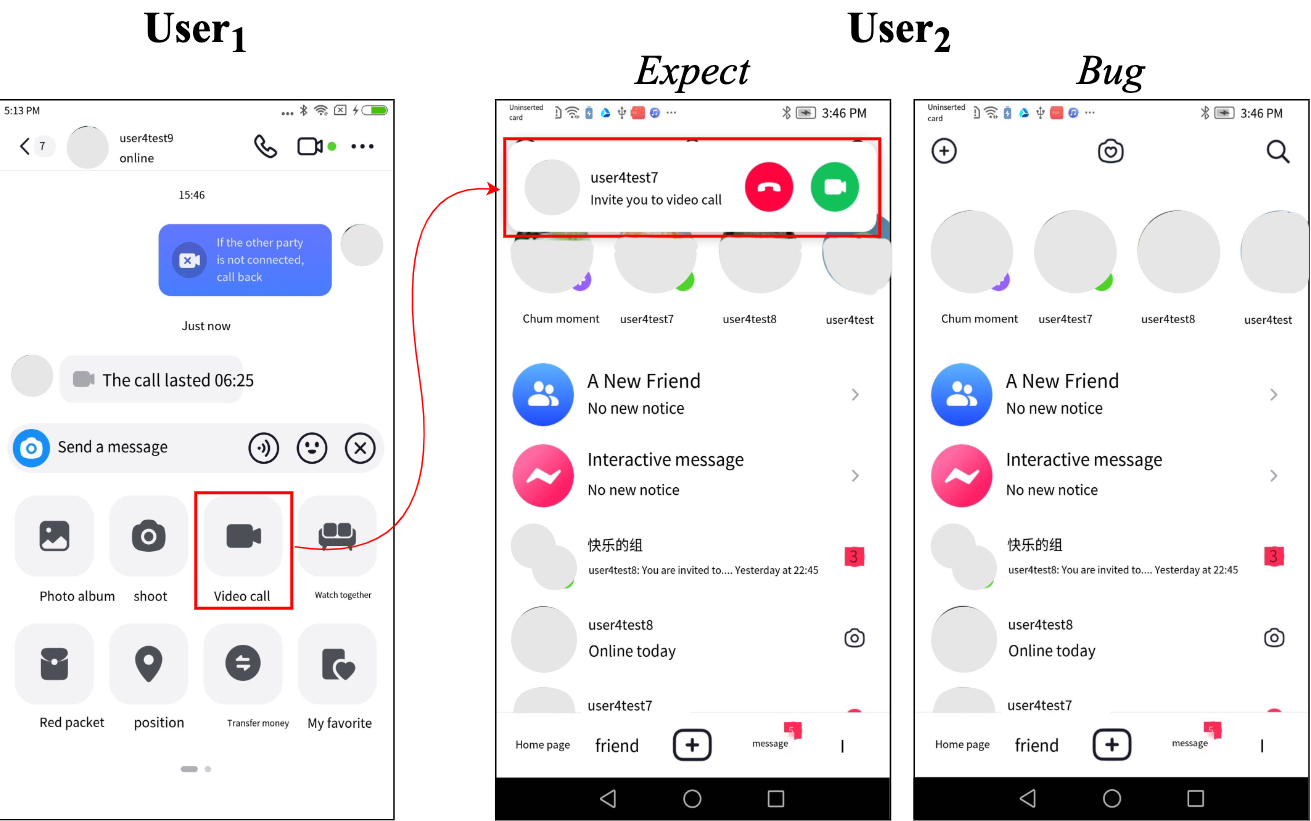}
	\caption{Illustration of interactive bug.}
	\label{fig:bug}
\end{figure}

\textbf{Results.} 
We run the experiment in their testing system, with over 3,318 task descriptions in one month from August 2024 to September 2024.
The completion rate for task automation reached 70.09\%, indicating that most task descriptions can be successfully automated and tested without the need for developer interaction. 
Of the instances that failed, we manually identify 26 bugs.
Fig.~\ref{fig:bug} illustrates an example of a common multi-user interactive bug where a user invites another to a call, but the user does not receive the call, hindering the automated process failure.
This indicates the potential value of our approach in detecting multi-user interactive bugs, which can significantly facilitate the overall app testing process in practice.

\section{Discussion}
\label{sec:discussion}
We discussed the limitations of our approach in automating the multi-user interactive tasks due to LLM inference time constraints, cross-app interaction, and ambiguous GUI design in Section~\ref{sec:rq1}.
In this section, we discuss the industrial implications and threats to the validity of our approach.

\subsection{Industrial Implication}
Before the implementation of the virtual device farm, testing for the TikTok app was exclusively performed on a physical device farm. 
This process, particularly for multi-user interactive feature testing, often requires a substantial number of physical devices from various platforms, SDKs, and vendors. 
Such testing incurs high operation costs and reduces device lifetime, hindering the adoption of modern rapid testing practices~\cite{savor2016continuous,wang2023test}  to enable continuous integration and deployment (CI/CD).
To address this, we introduce a virtual device farm designed to emulate a diverse range of mobile devices. 
We identify this solution as a preliminary stage in our app testing process, with the virtual device farm providing the majority of the testing workload, such as scalable device emulation for multi-user interaction and the identification of device-agnostic bugs.
Subsequently, the physical device farm acts as a last-level defense, focusing on uncovering hardware-specific issues.
This two-tiered approach optimizes the testing process by reducing costs and enhancing effectiveness and efficiency.

Task automation aims to cover the test of specific objectives, in order to reduce the time developers spend on extensive code scripting, e.g., on average, saving 87\% of the time in Section~\ref{sec:rq2}.
The ultimate goal of task automation is to facilitate the automatic identification of bugs, thereby enhancing efficiency and productivity in the development cycle.
Despite the availability of numerous advanced bug detection tools~\cite{web:monkey,li2019humanoid,chen2023unveiling}, we have opted for manual bug detection in the industrial app TikTok, informed by our practical insights.
That is, automated tools face difficulties in identifying interactive bugs that may not manifest on a single device but become apparent during multi-device interactions. 
For example, as depicted in Fig.~\ref{fig:bug}, a call-receiving bug is unable to automatically be identified by the existing bug detection tools, due to their limitations in detecting multi-device interaction issues.


Another potential implication is the applicability of our approach to a broader range of apps, such as social media apps like Facebook and Twitter, where multi-user interaction is a core component.
Results in Section~\ref{sec:evaluation} have initially demonstrated the effectiveness, usefulness, and practicality of our approach when employing specific multi-user interactive tasks from TikTok in real-world practice.
Although our study focuses on the TikTok app, we anticipate that our approach can be easily adapted to other apps.
The reason is that the foundational principles of our approach focus on the interaction patterns and LLM-based agent designs, which are not specific to any single app.

\subsection{Threats To Validity}
In our experiments to evaluate our approach, potential threats to internal validity may emerge due to the data leakage issues from LLMs.
To mitigate this bias, two authors independently undertook the task of manually labeling new tasks specifically curated for this research. 
These tasks comprise the experimental dataset used to evaluate our approach.
Subsequently, a threat to validity may arise from the quality of these annotated new tasks. 
To reduce the bias of subjectivity or mistakes during this labeling process, the authors carried out the annotation process independently and without prior discussion. 
A following consensus discussion was held to finalize the dataset. 
Moreover, to enhance the validity of our tasks in a real-world context, we engaged professional internal developers to further review the annotated tasks.

Another threat to validity may arise from the representativeness of our experiment due to the inherent randomness in LLM output generation, which could yield varying results across different executions.
Specifically, the metrics obtained from identical prompts might fluctuate between runs. 
To mitigate this threat, we conduct all the results across three separate runs of the LLMs-related methods (i.e., our approach and the baselines) to address the potential variability of the LLMs.




\section{Related Work}
Our work leverages a multi-agent framework to enhance the testing of multi-user interactive features within the app. Therefore, we discuss the related work, including automated app testing and multi-agent software testing. 

\subsection{Automated App Testing}

A growing body of tools has been dedicated to assisting in automated app testing, based on randomness/evolution~\cite{mao2016sapienz,ye2013droidfuzzer,web:monkey}, UI modeling~\cite{choi2013guided,gu2019practical,xie2022psychologically,xie2020uied}, systematic exploration~\cite{anand2012automated,feng2022gallery,chen2019gallery}, and LLMs~\cite{liu2023make,wang2024feedback,feng2024mud}.
However, these tools are often constrained to test activities that are triggered by specific features.
As opposed, numerous studies focus on creating tests that target specific features, directed by manually written descriptions. 
This approach is most prominently embodied in the field of script-based record and replay techniques, such as RERAN~\cite{gomez2013reran}, WeReplay~\cite{feng2023towards,feng2023efficiency}, and Sikuli~\cite{yeh2009sikuli}.
However, the reliance of these scripts on the absolute positioning of GUI elements or on brittle matching rules poses significant challenges to their adoption, as these factors can lead to a lack of robustness in automated testing.

In an effort to improve upon existing methods, many studies~\cite{linares2017developers,feng2024prompting,hu2018appflow} aim to employ natural language descriptions that delineate the target test features.
For instance, CAT~\cite{feng2024enabling} introduces a novel approach based on Retrieval Augmented Generation (RAG) techniques to generate specific tests (e.g., share a picture).
However, these efforts typically focus on single-user features and often overlook multi-user interactive features, which require several users to collaborate dynamically to trigger the feature. 
This oversight can impede achieving high test coverage. 
In our work, we propose a novel approach that utilizes multi-agent Large Language Models to mock up several users in testing multi-user interactive features, thereby enhancing the effectiveness of the app testing process.




\subsection{Multi-agent Software Testing}

Agents have been increasingly applied to support the automation of different software testing activities. 
One such approach relates to test management~\cite{malz2010agent,arora2018systematic}, which aims at selecting an appropriate set of test cases to be executed in every software test cycle using test unit agent and fuzzy logic.
Nevertheless, the testing of certain complex software features, such as chat functions, multi-player gaming, and interactions between users with varying levels of access, cannot be effectively conducted by a single-agent.
To address this limitation, a body of research~\cite{kumaresen2020agent,ramirez2021agent,enoiu2019test} has explored the concept of multi-agent-oriented software testing.
This involves the creation of multiple intelligent agents specifically designed to tackle the testing of these intricate features.
For instance, Tang~\cite{tang2010towards} introduced a variety of agents, each specialized in different roles equipped with hard-coded capabilities, including test design, execution, and evaluation.
Similarly, Dhavachelvan et al.~\cite{dhavachelvan2006new,dhavachelvan2005multi} have crafted several ad-hoc agents with predefined patterns to test specific features within the software.
However, these approaches rely on hard-coded rules tailored to individual software, which poses challenges for generalization across different software.

Large Language Models (LLMs) have been harnessed to facilitate a wide array of software engineering tasks.
A notable insight is that LLMs, trained on extensive datasets, can produce a compelling sense of being in the presence of a human-like interlocutor, denoted as an autonomous agent, capable of dynamically reacting to contextual cues.
Inspired by this capability, our work incorporates a multi-agent framework that integrates LLMs to address the complexities involved in testing multi-user interactive software features. 
That is, we deploy a set of autonomous agents designed to simulate multiple users.
Each agent is responsible for interacting with the software environment, with the goal of fulfilling their individual assigned tasks. 
Together, the agents collaboratively undertake a comprehensive test of the software's multi-user interactive features.

\section{Conclusion}
Multi-user interactive features are a critical component of the software, particularly in the TikTok app.
To address the challenges associated with testing these features, we introduce a novel multi-agent approach designed to automatically trigger multi-user interactive features.
In detail, our approach utilizes a series of agents, each driven by Large Language Models (LLMs), to perform autonomous actions that collectively contribute to the automation of interactive tasks on devices.
The experiments demonstrate the effectiveness and usefulness of our approach in automatically executing the multi-user interactive tasks in the TikTok app.
Moreover, our approach has been successfully integrated into the industrial TikTok testing platform, enhancing 3,318 interactive feature testing and improving 26 bug detection.
This integration underscores the practical value and impact of our approach in real-world software testing scenarios.

In the future, we intend to refine our approach in two aspects.
First, considering that multi-user interactive features are often time-sensitive, we aim to boost the efficiency of our approach by implementing open-source LLMs on the local server, hence reducing latency and improving response times. 
Second, interactive features can span multiple applications, for instance, the interaction of sharing a video stream requires seamless interactivity between the current service and the messaging-sharing platform.
We will explore the potential of our approach to support multi-user cross-app interactions.

\section*{Acknowledgment}
We appreciate the assistance from Mengfei Wang and the Douyin team for their valuable contributions to the methodology discussion and experimental processes. 

\bibliographystyle{IEEEtran}
	\bibliography{main}
\end{document}